# Valley-Spin Hall Effect-based Nonvolatile Memory with Exchange-Coupling-Enabled Electrical Isolation of Read and Write Paths

Karam Cho, and Sumeet Kumar Gupta, *Senior Member, IEEE*

*Abstract*—Valley-spin hall (VSH) effect in monolayer $WSe_2$ has been shown to exhibit highly beneficial features for nonvolatile memory (NVM) design. Key advantages of VSH-based magnetic random-access memory (VSH-MRAM) over spin orbit torque (SOT)-MRAM include access transistor-less compact bit-cell and low power switching of perpendicular magnetic anisotropy (PMA) magnets. Nevertheless, large device resistance in the read path ($R_S$) due to low mobility of $WSe_2$ and Schottky contacts deteriorates sense margin, offsetting the benefits of VSH-MRAM. To address this limitation, we propose another flavor of VSH-based MRAM that (while inheriting most of the benefits of VSH-MRAM) achieves lower $R_S$ in the read path by electrically isolating the read and write terminals. This is enabled by coupling VSH with electrically-isolated but magnetically-coupled PMA magnets via interlayer exchange-coupling. Designing the proposed devices using object oriented micro magnetic framework (OOMMF) simulation, we ensure the robustness of the exchange-coupled PMA system under process variations. To maintain a compact memory footprint, we share the read access transistor across multiple bit-cells. Compared to the existing VSH-MRAMs, our design achieves 39%-42% and 36%-46% reduction in read time and energy, respectively, along with 1.1X-1.3X larger sense margin at a comparable area. This comes at the cost of 1.7X and 2.0X increase in write time and energy, respectively. Thus, the proposed design is suitable for applications in which reads are more dominant than writes.

## I. INTRODUCTION

WITH the recent explosive growth in highly data-centric applications such as internet of things and artificial intelligence, high storage capacity along with low-power computing in memories has become more crucial than ever. Although conventional CMOS-based memories have brought a great prosperity to the semiconductor industry to date enabling high-performance computing, increasing leakage energy and low cell density hinder them from sustaining their benefits in the deep submicron regime [1]. To tackle such issues, emerging nonvolatile memories (NVMs) such as phase change memory (PCM), resistive random-access memory (RRAM), ferroelectric (FE) memories and spin-based magnetic RAMs (MRAMs) are being extensively explored [1].

Among the emerging NVMs, spin transfer torque (STT)-MRAM has demonstrated the highest endurance with comparable performance (> $10^{12}$ cycles of endurance and < 10 ns write time), is CMOS compatible [1], [2] and hence, is considered to be one of the most promising NVMs [3]. In addition to the fast operation, high endurance and non-volatility, it also exhibits distinct advantages over CMOS memories such as low standby power and high density. However, its high write current density flowing through a magnetic tunnel junction (MTJ) causes critical design conflicts leading to high write power and time-dependent breakdown of the tunneling oxide in the MTJ. Moreover, low distinguishability between two binary states (measured by tunneling magneto-resistance, or TMR) aggravates the design challenges, putting questions on the wider adoptability of STT-MRAM in advanced applications [2], [3].

To address some of the issues of STT-MRAM, spin orbit torque (SOT)-MRAM has been proposed following the discovery of giant spin hall (GSH) effect in heavy metals (e.g. Pt, Ta, and W) [4]–[6]. SOT-MRAM (also named 'GSH-MRAM') has separate write and read paths (Fig. 1(a)). Write path through a metal leads to substantial reduction in the write power. Besides, the separation between write and read paths in SOT-MRAM mitigates the reliability issues [7]. However, these benefits come at the cost of higher area due to the need for multiple access transistors per bit-cell (see Fig. 1(b) and (c)) [7], [8]. A technique based on sharing of the read access transistor alleviates some area overhead [9], but has some write issues as the read and write paths are not *completely* electrically isolated. Another limitation of GSH-MRAM is that its field-free operation relies on in-plane magnetic anisotropy (IMA) which is inferior to perpendicular magnetic anisotropy (PMA) in the context of write efficiency. Coupling GSH effect with PMA typically requires an external magnetic field leading to design overheads [10]. Some schemes have been introduced to avert an external magnetic field such as interplay of STT and SOT [11], [12], structural asymmetry [13], and interlayer exchange coupling [14]. However, such designs have their own

Manuscript received September 16, 2022. This work was supported by SRC/NIST-funded NEWLIMITS Center (Award number 70NANB17H041).

Karam Cho, and Sumeet Kumar Gupta are with the School of Electrical and Computer Engineering, Purdue University, West Lafayette, IN 47907 USA (e-mail: cho346@purdue.edu; guptask@purdue.edu).

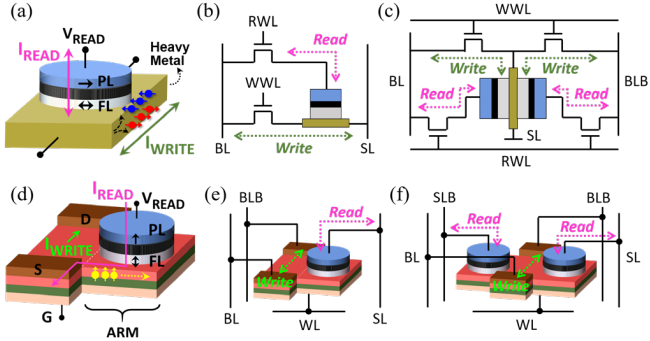

Fig. 1. (a) SOT(GSH)-MRAM and (d) VSH-MRAM [16] and their array-level architectures for (b, e) single-ended and (c, f) differential memories.

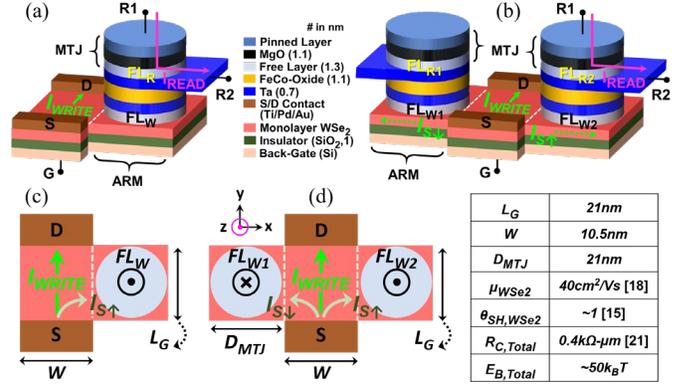

Fig. 2. Structure of the proposed (a) single-ended EIRW-VSH and (b) differential DEIRW-VSH devices where the electrical isolation of read and write paths is achieved via free layers ($FL_W$ and $FL_R$) which are magnetically-coupled but electrically-isolated. The illustration of write operation in (c) and (d) shows that $I_{WRITE}$-induced spin currents with opposite out-of-plane spins ($I_{S\uparrow}/I_{S\downarrow}$) switch $FL_W$ upward/downward.

overheads. For example, the design in [14] still relies on IMA based design, and the read path is not completely isolated from the write path.

Valley spin hall (VSH) effect in 2D transition metal dichalcogenides (TMDs) can potentially be a key for implementing an energy-efficient PMA-based spintronic memory [16]. VSH effect enables field-free PMA switching even with higher spin injection efficiency with valley hall angle ($\theta_{SH}$) as large as 1 at room temperature [15], significantly reducing the write power. Additionally, as we previously proposed in [16], the VSH-MRAM in [16] (Fig. 1(d)) utilizes an integrated back-gate to control the charge and spin currents in 2D TMD channel. This leads to an access transistor-less compact bit-cell (Fig. 1(e) and (f)). However, large channel resistance due to low mobility of monolayer WSe$_2$ and source/drain Schottky contacts results in large series resistance ($R_S$) in the read path, deteriorating the memory sense margin.

Motivated by this, we propose VSH-based spintronic memories that not only inherit the appealing features of existing VSH-MRAM but also attain lower $R_S$ by electrically isolating the read and write paths (EIRW). We name the proposed memories EIRW-VSH MRAMs. Low $R_S$ is achieved by designing electrically-isolated but magnetically-coupled PMA magnets via interlayer exchange-coupling [17]. Previously, we had explored similar structure in the context of nonvolatile flip-flop design [18]; here, we utilize EIRW-VSH for memory design by appropriately optimizing the device for nonvolatile memory applications and carrying out device-array co-design to achieve optimal energy-performance-area trade-offs. The key contributions of this paper are as follows:

- We propose single-ended and differential nonvolatile spintronic memories utilizing VSH effect in monolayer WSe$_2$, which feature 1) out-of-plane spin generation that can be coupled with PMA magnets, 2) integrated back-gate for spin current control during write, 3) electrical isolation of read and write paths via exchange-coupling between PMA magnets and 4) shared read access transistor for compact array design.
- We investigate the impact of process variations on the exchange coupling between PMA magnets and the functionality of the proposed memories.
- We explore the implications of shared read access transistor for both single-ended and differential EIRW-VSH MRAMs, advancing the study in [9] (which focused on single-ended memories).
- We analyze the benefits and trade-offs of the electrical isolation of read and write paths at device- and array-levels, and conduct a detailed comparison with previously proposed VSH-MRAM.
- We benchmark memory characteristics with SRAM and emerging spin-based nonvolatile memories.

II. VSH-Devices With Exchange-Coupling-Enabled Electrical Isolation of Read and Write Paths (EIRW)

A. Structure and Operation

As proposed in our earlier work [18], the electrical isolation of read and write paths (EIRW) in VSH-devices is achieved by utilizing the exchange-coupling between PMA free layers (FLs). In contrast to [18] where the device was designed with small energy barrier ($E_B$) of FLs (suitable for nonvolatile flip-flops), here, the exchange-coupled PMA FLs are optimized to have $E_B$ of ~50 $k_B$T to ensure large memory lifetime (> 10 years). Moreover, since low device footprint is of paramount importance in memory design, we optimize the FLs to have smaller dimension compared to [18] while maintaining ~50 $k_B$T energy barrier. For this, we use a different set of experimental parameters [19] to obtain uniaxial anisotropy density ($K_U$) and saturation magnetization ($M_S$) which are suitable for memory design with larger $E_B$ (see Table I). Employing these parameters, we propose and optimize two flavors of memory devices: single-ended EIRW-VSH and differential DEIRW-VSH (see Fig. 2(a) and (b), respectively).

The proposed EIRW-VSH devices are designed to exploit the VSH effect in monolayer WSe$_2$. They include a spin generator, which has a p-type monolayer WSe$_2$ channel and one or two arms. The device is designed with a single arm (Fig. 2(a)) for single-ended read functionality (EIRW-VSH) or with two arms (Fig. 2(b)) for a differential read (DEIRW-VSH). WSe$_2$ channel and arms have an integrated back-gate that controls the flow of charge current ($I_C$) in the channel and in turn, spin currents ($I_S$) in the arms (which are generated due to the VSH effect). The $I_S$ interacts with free layers, $FL_W$s, which are in direct contact with WSe$_2$. $FL_W$ on each arm is a 1.3-nm CoFeB layer with perpendicular magnetic easy axis in the z-direction. A structure of Ta (0.7 nm)/FeCo-oxide (1.1 nm)/Ta (0.7 nm)/CoFeB (1.3 nm) is formed on top of $FL_W$ to ferromagnetically couple $FL_W$ with the top CoFeB FL ($FL_R$) via interlayer exchange-coupling [17]. An MTJ is formed on top, having $FL_R$ as its FL. A *Read* path is formed between nodes R1 and R2 (through MTJ) by extending the upper Ta layer in the stack. Note, the read path is completely electrically isolated from the WSe$_2$ channel (or *Write* path). Therefore, the proposed devices achieve an electrical isolation of read and write paths (EIRW) utilizing the exchange-coupling between PMA FLs (i.e. $FL_W$ and $FL_R$). In terms of conducting properties, thin film Ta has lower resistivity (2 KΩ-nm [20]) than monolayer WSe$_2$ thus leading to low $R_S$ in our designs compared to VSH-MRAMs [16].

A *Write* path is formed between source (S) and drain (D) contacts. When $I_C$ (or $I_{WRITE}$) flows between S and D in the *y* direction (see Fig. 2), spin currents with opposite out-of-plane spin polarization ($I_{S↑}/I_{S↓}$) are induced to flow along *x*-axis. Induced $I_{S↑}$ ($I_{S↓}$) switches $FL_W$ upward (downward) by exerting spin torque. Note, EIRW-VSH device holds either $I_{S↑}$ or $I_{S↓}$ along the arm. On the other hand, DEIRW-VSH device has both $I_{S↑}$ and $I_{S↓}$ in the two arms, leading to complementary storage of the bit information in the two $FL_W$s (see Fig. 2(c) and (d), respectively). Then, $FL_R$ is switched by interlayer exchange fields generated from the corresponding $FL_W$. Thus, $I_C$ in WSe$_2$ leads to $I_{S↑}/I_{S↓}$, which, in turn, set the MTJ state to be in P- or AP-state via the combination of spin torque and interlayer exchange coupling. The state of the MTJ depends on the direction of the current ($I_C$ or $I_{WRITE}$): In case of DEIRW-VSH device, when $I_C$ flows in +*y* direction, $I_{S↑}$ ($I_{S↓}$) is generated in +*x* (-*x*) direction, and $FL_{W2}$ ($FL_{W1}$) switches upward (downward). This switches $FL_{R2}$ upward and $FL_{R1}$ downward via exchange coupling, encoding P and AP states in MTJs on the right and left arms, respectively (the magnetization of PL is in $+\hat{z}$). In EIRW-VSH device, $I_C$ flowing in +*y* direction induces $I_{S↑}$ in +*x* direction, setting the MTJ to be in P-state. The opposite happens if the $I_C$ direction is reversed.

### B. Simulation Framework

A simulation framework (Fig. 3) is established to evaluate the proposed EIRW-VSH devices and arrays, and to compare with VSH-MRAM based designs [16]. First, we build a 2D FET model that self-consistently captures 2D electrostatics and charge transport in p-type WSe$_2$ channel. The drain current (or charge current, $I_C$) is calculated using the extracted mobility of monolayer WSe$_2$ and contact resistance ($R_C$) at S/D side [21]. Calibration with experiment of the drain current is in inset of 2D FET model (details on experiment of WSe$_2$ FET in [18]). Since both VSH-MRAMs and EIRW-VSH devices exploit the VSH effect in WSe$_2$, the 2D FET model is used for all designs.

For the proposed EIRW-VSH devices, $I_C$ (as a function of $V_{GS}$ and $V_{DS}$ of the device), is delivered to micromagnetic OOMMF simulation (Object Oriented Micro Magnetic Framework [22]) to investigate the implication of current-induced SOT on the coupled PMA FLs (recall the '*write*' operation). To capture the VSH effect-driven spin torque, we use an existing 'Xf_STT' class in OOMMF and modify the equations of STT [23] for our memory device structure (dimensions in Fig. 2 and Table I). We also use a 'TwoSurfaceExchange' class to integrate the RKKY-style exchange-coupling (coupling strength, $J_{EX}$) between PMA FLs ($FL_W$ and $FL_R$). We use $J_{EX} = 0.35$ mJ/m$^2$ as reported in the experiments in [17]. The uniaxial anisotropy ($K_U$) and exchange stiffness ($A$) which are related to the intra-exchange energy of PMA FLs, are considered. For OOMMF simulation, we choose cell sizes (or mesh sizes) as 1nm, 1nm, and 1.3nm (x, y, and z) and the magnetic parameters used in simulation are listed in Table I. The switching time of the coupled PMA system (as the result of VSH) obtained from OOMMF simulation is provided to HSPICE simulation for array-level analysis. The MTJ resistance which is involved in the read operation (details in section III) is obtained from NEGF equations, as detailed in [24], and used to develop a Verilog-A model for HSPICE simulations. For baseline VSH-MRAMs, we follow the simulation process outlined in [16].

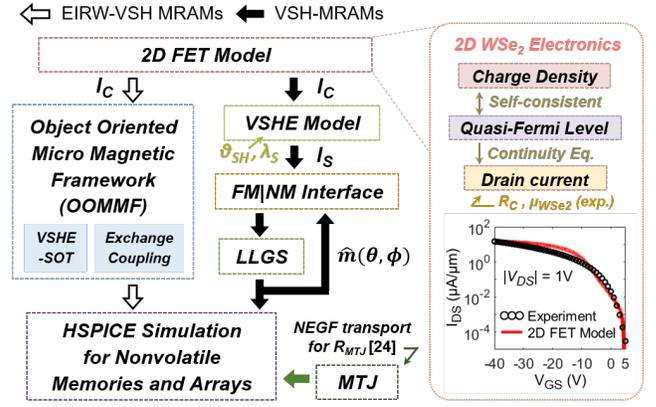

Fig. 3. Simulation framework for evaluation and comparison of the proposed EIRW-VSH memory devices with VSH-MRAMs [16]. Calibration of monolayer 2D WSe$_2$ FET with experiment [18] is in the extended inset of 2D FET Model.

TABLE I
PARAMETERS FOR MICROMAGNETIC (OOMMF) AND HSPICE SIMULATION

| Parameter | VSH-MRAM | EIRW-VSH MRAM |
|---|---|---|
| Thickness of Single FL [nm] | 1.3 | 1.3 |
| Volume of Single FL [nm$^3$] | (π/4)×30×30×1.3 | (π/4)×21×21×1.3 |
| Thickness of MgO [nm] | 1.2 | 1.1 |
| Saturation Magnetization, $M_S$ [emu/cm$^3$] | 1257.3 [19] | 1257.3 [19] |
| Uniaxial Anisotropy Density, $K_U$ [erg/cm$^3$] | 2.3×10$^6$ [19] | 2.3×10$^6$ [19] |
| Damping Coefficient, α | 0.008 [25] | 0.008 [25] |
| Gyromagnetic Ratio, γ [MHz/Oe] | 17.6 | 17.6 |
| Energy Barrier, $E_B$ [k$_B$T] | ~50 | ~50 |
| Exchange Stiffness, $A$ [pJ/m] | 13 | 13 |
| Exchange Coupling Strength, $J_{EX}$ [mJ/m$^2$] | - | 0.35 [17] |
| Resistivity of Ta [KΩ-nm] | - | 2 [20] |

### C. Device Characteristics

As mentioned earlier, the proposed (D)EIRW-VSH devices are re-designed from our previous work [18] for evaluation and comparison with (D)VSH-MRAMs [16]. We perform iso-thermal stability ($\Delta = E_B/k_BT$) analysis between the designs. In the baseline (D)VSH-MRAMs, the single FL is designed to have $\Delta$~50. On the other hand, each FL in (D)EIRW-VSH devices is designed to have smaller $\Delta$ but $\Delta$~50 as a whole system (i.e. exchange-coupled $FL_W$ and $FL_R$). This is achieved by maintaining the same thickness but different diameter of a single FL in two designs (see Table I for details). This results in smaller area footprint in our proposed designs compared to (D)VSH-MRAMs, which will be discussed in section III. The thickness of MTJ oxide (i.e. MgO) is optimized separately for each device to maximize the SM. Specifically, due to the lower $R_S$ in (D)EIRW-VSH devices, the optimal MgO thickness is lower than (D)VSH devices (Table I).

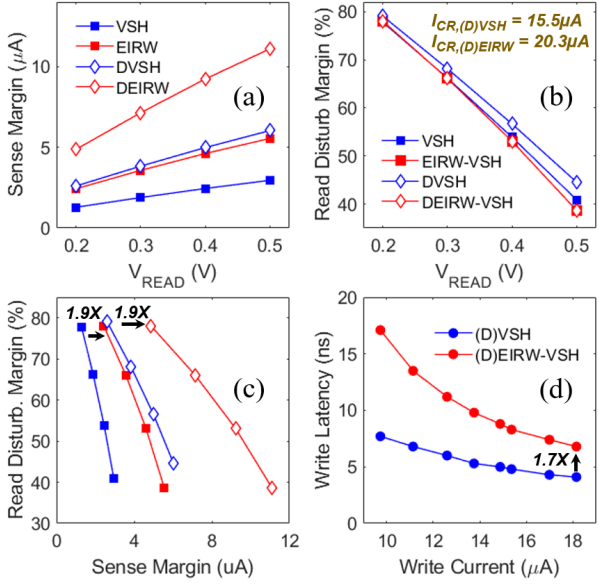

Fig. 4. Device-level analysis of (D)EIRW-VSH devices compared with (D)VSH-MRAMs: (a) sense margin (SM) and (b) read disturb margin (RDM) versus read voltage ($V_{READ}$), (c) 1.9X enhancement in SM is achieved at ~iso-RDM of ~80%, (d) write latency versus write current.

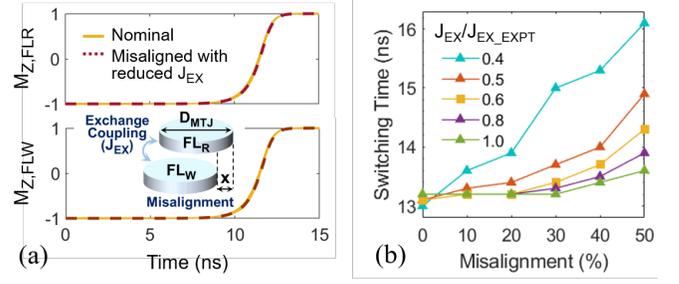

Fig. 5. (a) Magnetization versus time of nominal scenario (solid) and exemplary case (dotted) with 20% misalignment and 40% reduction in exchange coupling strength ($J_{EX}$). Inset: misalignment (= $x/D_{MTJ}$ *100) between $FL_W$ and $FL_R$. (b) Switching time (in ns) with various $J_{EX}$ and misalignment.

*1) Read*

For read, the exchange-coupling-enabled electrical isolation between read and write paths in (D)EIRW-VSH devices is favorable because of the following two aspects:

(1) The high $R_S$ along the $WSe_2$ channel found in the (D)VSH-MRAMs is averted in (D)EIRW-VSH devices, resulting in higher sense margin (SM; Fig. 4(a)). For single-ended designs (i.e. VSH-MRAM and EIRW-VSH device), SM is calculated as $(I_P - I_{AP})/2$ where $I_P$ and $I_{AP}$ denote $I_{READ}$ flowing through P- and AP-state MTJs, respectively. For differential designs (i.e. DVSH-MRAM and DEIRW-VSH device), SM is calculated as $I_P - I_{AP}$. Due to the low $R_S$ in the read path and thus larger $I_{READ}$, (D)EIRW-VSH devices exhibit 1.9X improved SM compared to (D)VSH-MRAMs in the entire range of $V_{READ}$ considered in Fig. 4(a).

(2) Second, the exchange-coupling between $FL_W$ and $FL_R$ in (D)EIRW-VSH devices can be beneficial for lowering the read disturbance. The disturbance is attributed to the fact that $I_{READ}$ flowing through an MTJ can exert STT and disturb the MTJ state (see Fig. 1(d) and Fig. 2). Given the read current direction, the AP-state of MTJ is vulnerable to this disturbance and can switch from AP to P state (for both the baseline and proposed designs). However, the critical current ($I_{CR}$) above which such undesired switching occurs is higher in (D)EIRW-VSH devices ($I_{CR}$ = 20.3μA) compared to (D)VSH-MRAMs ($I_{CR}$ = 15.5μA). This is because $FL_W$ generates an effective field (due to exchange coupling) which opposes the switching of $FL_R$ and helps in lowering the disturbance due to $I_{READ}$. However, as the $I_{READ}$ is also higher due to the low $R_S$ in (D)EIRW-VSH devices, read disturb margin (RDM), which is defined as ($I_{CR}-I_{AP})/I_{CR}\times100$, is *slightly* lower for the proposed devices at the same $V_{READ}$ (Fig. 4(b)). This reduction in RDM is much smaller compared to the SM improvements discussed earlier. Therefore, at ~iso-RDM of ~80% (Fig. 4(c)), it is observed that SM in (D)EIRW-VSH devices is 1.9X larger.

*2) Write*

Figure 4(d) shows the write latency *versus* $I_{WRITE}$ of (D)VSH-MRAMs and (D)EIRW-VSH devices. As expected, increase in $I_{WRITE}$ facilitates the faster switching of FLs in both designs. However, due to the exchange-coupling in (D)EIRW-VSH devices, write latency increases >1.7X compared to (D)VSH-MRAMs for the same write current. This is the cost of introducing electrical isolation of read-write paths, which is primarily aimed at improving the read operation, as discussed above.

*D. Impact of Process Variations on Exchange Coupling*

We also examine the implication of potential process variations on write performance (i.e. switching time of the exchange-coupled PMA magnets). Albeit such study was performed in our earlier work in the context of nonvolatile flip-flops [18], the FLs re-designed for memories with a higher $E_B$ in this paper require a separate analysis to understand their resilience to possible process variations. As shown in the inset in Fig. 5(a), we consider horizontal misalignment between $FL_W$ and $FL_R$, and reduction in exchange coupling strength ($J_{EX}$) as the process variation factors. The misalignment is swept from 0 to 50% as to the diameter of FL ($D_{MTJ}$), and the exchange coupling strength is reduced to 40%~100% of its original value ($J_{EX\_EXPT}$) reported in the experiment [17]. Here, 0% of misalignment and 100% of $J_{EX\_EXPT}$ (i.e. $J_{EX} = J_{EX\_EXPT}$) is the nominal scenario. Even with misalignment as large as 20% and reduction in $J_{EX\_EXPT}$ by 40% (i.e. $J_{EX} = 0.6*J_{EX\_EXPT}$; Fig. 5(b)), we observe a successful switching, with the switching time of the exchange-coupled PMA system remaining similar to the nominal case (Fig. 5(a)). Moreover, reduction in $J_{EX\_EXPT}$ by 50% (i.e. $J_{EX} = 0.5*J_{EX\_EXPT}$) leads to only 1.5% increase in switching time for 20% misalignment. For larger misalignment values, which are unlikely to occur, we observe that switching of $FL_R$ via exchange coupling is still attainable as long as $J_{EX}$ remains greater than 40% of the experimental value. However, we notice switching failures in $FL_R$ (when $FL_W$ reversed by VSH effect) for reduction in $J_{EX\_EXPT}$ by 70% (i.e. $J_{EX} = 0.3*J_{EX\_EXPT}$). Therefore, we conclude that exchange-coupled PMA FLs are resilient to the potential process variations without significantly degenerating its performance keeping large margins in both the misalignment and $J_{EX}$ variation.

## III. (D)EIRW-VSH Device Based Memory Arrays

### A. Design of Memory Arrays based on (D)EIRW-VSH Devices with Shared Read Access Transistor

Utilizing the proposed (D)EIRW-VSH devices, we design single-ended and differential memory arrays (Fig. 6) with 256 rows and 256 columns. As the read path in the proposed (D)EIRW-VSH devices is formed along the MTJ and Ta layer, a read access transistor is required to control the access to the designated bit-cells during read and to avoid the sneak current through the unaccessed cells. However, having the read access transistor per bit-cell can lead to the large area penalty in our proposed designs compared to the (D)VSH-MRAMs. Therefore, for array design, we adopt a scheme proposed previously by us in [9] for GSH-MRAMs where bit-cells belonging to the same word share the read access transistor as shown in Fig. 6. Note, in [9], the read and write paths are not completely isolated, which can cause write issues/overheads in GSH-MRAMs. However, due to the complete electrical isolation of the read and write paths in our proposed devices such issues are eliminated, and the read path can be optimized solely considering the read requirements. Moreover, the analysis in [9] only considers single-ended design. In this work, we also investigate the sharing of the read access transistor for a differential design. Specifically, for DEIRW-VSH MRAM, the read access transistor is connected to *both the true and complementary* MTJs of the bit-cells belonging to a word.

During read, we access 64 bit-cells (considering a 64-bit word) by asserting the shared read access transistor. Therefore, the read access transistor is shared with 64 MTJs for EIRW-MRAMs and with 128 MTJs for DEIRW-MRAMs. The access transistor is an n-type FinFET with its gate connected to read-word-line (RWL). Node R1 of each bit-cell (see Fig. 2) is connected to the drain of the shared read access transistor. Sense-lines, SL (and SLB) are connected to node R2 along the extended Ta layer in the MTJ stack of (D)EIRW-VSH devices and are shared along the column. With this arrangement, read current ($I_{READ}$) through each bit-cell can be read individually on SL (and SLB). However, sharing the read access transistor in a word alters the read operation as the MTJs of 64 bit-cells are connected in parallel, and the total resistance affects current flowing through the shared read access transistor (details will be discussed shortly). The increase in area due to the shared read access transistor is also discussed later in this section.

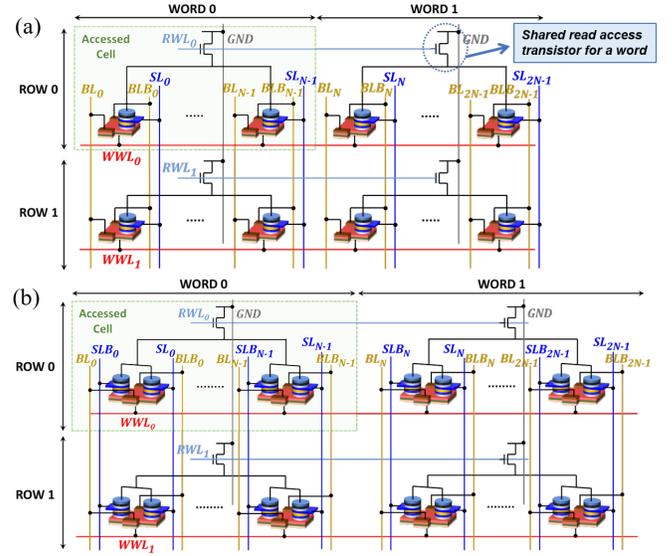

Fig. 6. Memory array architectures with bit-cells belonging to the same word sharing the read access transistor [9] for the proposed (a) EIRW-VSH MRAM and (b) DEIRW-VSH MRAM.

TABLE II
OPERATING BIAS CONDITIONS WHEN WORD 0 OF ROW 0 IS ACCESSED

|  |  | Write | Read |
|---|---|---|---|
| Accessed | WWL$_0$ | 0 | V$_{DD}$ |
|  | BL$_{0:N-1}$ | V$_{DD}$/0 | 0 |
|  | BLB$_{0:N-1}$ | 0/V$_{DD}$ | 0 |
|  | RWL$_0$ | 0 | V$_{DD}$ |
|  | SL(B)$_{0:N-1}$ | 0 | V$_{READ}$ |
| Unaccessed | WWL$_1$ | V$_{DD}$ | V$_{DD}$ |
|  | BL$_{N:2N-1}$ | V$_{DD}$ | 0 |
|  | BLB$_{N:2N-1}$ | V$_{DD}$ | 0 |
|  | RWL$_1$ | 0 | 0 |
|  | SL(B)$_{N:2N-1}$ | 0 | 0 |

Similar to [16], an access transistor in the write path is not required in our array designs (see Fig. 6). This is because the back-gate of each bit-cell (i.e. (D)EIRW-VSH MRAMs) has the control of ON and OFF states of the (D)EIRW-VSH devices (see details in section II and in [16]) and thus, can be directly connected to write-word-line (WWL). The source and drain of each cell are connected to bit-line (BL) and bit-line-bar (BLB), respectively, which are shared along the cells in the same column. The details on operation are discussed in the next sub-section with bias conditions presented in Table II.

### B. (D)EIRW-VSH MRAM Memory Array Operation
#### 1) Write

To write, we access the selected bit-cell by applying 0V to the corresponding WWL (recall that the proposed memory device is p-type). Then, complementary biasing (i.e. $V_{DD}$/0) is applied to BL and BLB to flow $I_{WRITE}$ in the bit-cell. Note, depending on the BL/BLB biasing, $I_{S\uparrow}/I_{S\downarrow}$ will flow and encode Boolean logic into the MTJs in each bit-cell, as discussed earlier. For the single-ended array (Fig. 6(a)), if BL and BLB are driven to $V_{DD}$ and 0, respectively, $I_{WRITE}$ flows from source to drain, generating $I_{S\uparrow}$ to the right arm (due to the VSH effect) and switching $FL_W$ upward (for more details, see Fig. 2). As $FL_W$ and $FL_R$ are magnetically coupled, $FL_R$ also switches upward, encoding 'P' in the MTJ (i.e. logic '1' in the bit-cell). Encoding 'AP' (or logic '0') also can be done by applying 0 and $V_{DD}$ to BL and BLB, respectively. For differential array (Fig. 6(b)), the write mechanism is similar except for that the differential encoding in one device is possible. This is because the generated $I_{S\uparrow}$ and $I_{S\downarrow}$ flow in the opposite direction (Fig. 2 for more details), and store 'P' and 'AP' in MTJs located in two arms of the DEIRW-VSH MRAM. It is noted that, to avoid the sneak current paths, WWL, BL, and BLB are driven to $V_{DD}$ for unaccessed cells (hold condition) [16].

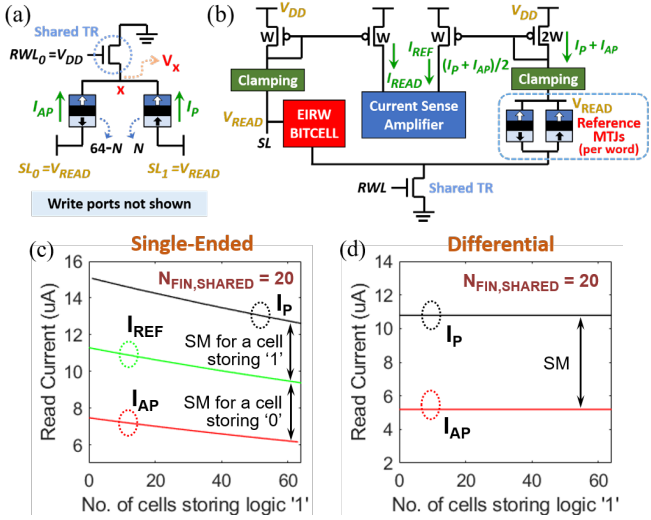

Fig. 7. (a) Equivalent circuit for read in a word (64-bit). (b) Reference circuit for current-based sensing where the reference MTJs share the read access transistor with 64 bit-cells [9]. Read current versus the number of cells storing logic '1' for (c) single-ended and (d) differential arrays.

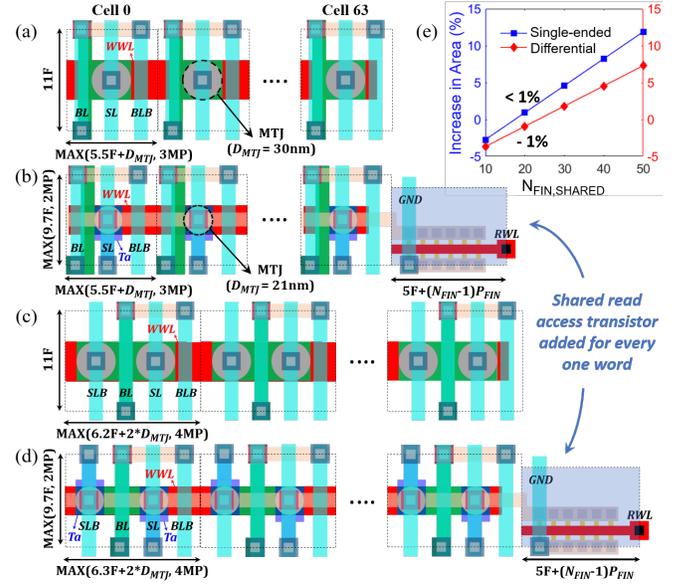

Fig. 8. Layout of a word (64-bit) for single-ended arrays with (a) VSH-MRAM and (b) EIRW-VSH MRAM, and differential arrays with (c) DVSH-MRAM and (d) DEIRW-VSH MRAM. Layout of the shared read access transistor added for every one word in (D)EIRW-VSH-MRAM arrays in (b) and (d), and the resultant area increase compared to (D)VSH-MRAM arrays with respect to the number of fins of the shared read access transistor ($N_{FIN,SHARED}$) in (e). $F$: minimum-feature size associated with a technology, MP: metal pitch, $D_{MTJ}$: MTJ diameter.

*2) Read*

To read the stored bits of an accessed cell, we enable the read path by applying $V_{DD}$ on RWL and $V_{READ}$ on SLs (and SLBs in case of DEIRW-VSH MRAM). We simultaneously read 64 bit-cells of the accessed word. Sharing the read access transistor among the bit-cells belonging to the same word has important read implications, especially in a single-ended array. Each MTJ in a bit-cell exhibits two levels of resistance, high- and low-resistance state ($R_{AP}$ and $R_P$, respectively), with $R_P < R_{AP}$. Therefore, when the 64 MTJs are connected to the read access transistor in parallel, the bit pattern stored, i.e. the number of cells storing logic '1/0' (P/AP-state), affects the equivalent resistance and controls the amount of currents flowing through the shared read access transistor and also, through the SL. To explain this, let us first consider a case where 'N' and '64-N' cells are storing logic '1' and '0', i.e. 'N' and '64-N' MTJs are in P- and AP-state, respectively (see Fig. 7(a)). The voltage at node x ($V_x$) is dictated by N. Now, if the number of cells storing logic '1' (N) increases, the equivalent resistance of all the MTJs decreases and thus, $V_x$ increases. This results in the decrease of voltage drop across the MTJs, reducing both $I_P$ and $I_{AP}$. Thus, $I_{READ}$ in the single-ended array is the function of the number of cells storing logic '1' as shown in Fig. 7(c).

Although we simultaneously read a word, each bit-cell can be sensed independently by reading $I_{READ}$ through each SL, which runs along the column and is connected to the current sense amplifier (CSA) in the reference circuit (Fig. 7(b)). Two reference MTJs (in P- and AP-state each) are employed in the reference circuit and also share the same read access transistor with the 64 bit-cells. By doing so, the reference current ($I_{REF} = (I_P + I_{AP})/2$ for single-ended array) flowing through the reference MTJs can track the number of cells storing logic '1' in the word (see Fig. 7(c)). The $I_{READ}$ (either $I_P$ or $I_{AP}$) from each bit-cell is compared to $I_{REF}$ and the output of the CSA yields the stored bit-information. Figure 7(c) shows $I_{READ}$ versus the number of cells storing logic '1' in single-ended array based on EIRW-VSH MRAM. As explained earlier, both $I_P$ and $I_{AP}$ decrease as the number of cells storing logic '1' increases. Here, SM of cells storing logic '1' and '0' are calculated as '$I_P - I_{REF}$' and '$I_{REF} - I_{AP}$,' respectively.

For the differential array designed with DEIRW-VSH MRAM, the number of cells storing '1' in a word is fixed to 64 regardless of the bit stored in each cell. This is because each differential cell holds both P- and AP-MTJs. Therefore, the $I_{READ}$ is constant irrespective of bit pattern stored in the accessed word (Fig. 7(d)). It is noted that in the case of the proposed differential array, the reference MTJs are not necessary as SL and SLB in a cell can be compared to determine the output (i.e. bit information of the cell). Here, SM is calculated as '$I_P - I_{AP}$.'

*C. Layout*

We also perform the layout analysis based on scalable $\lambda$-based rules [26], where $\lambda$ is half the minimum-feature size ($F$) associated with a technology. We consider gate and metal pitches corresponding to 7nm technology [27]. Figure 8 shows the layout comparison for a word. As shown in Fig. 8(b) and (d), the proposed arrays include the area of the shared read access transistor, leading to an increase in total width compared to (D)VSH-MRAM arrays. However, due to the smaller MTJ dimension (recall from Section II and Table I), bit-cell height is smaller in the proposed arrays, compensating the increase in width. Therefore, the increase in area (see Fig. 8(e)) in single-ended EIRW-VSH MRAM array is less than 1% compared to the VSH-MRAM array for $N_{FIN,SHARED}$ of 20. For differential arrays, DEIRW-VSH MRAM array exhibits 1% reduction in area compared to the DVSH-MRAM array at the same $N_{FIN,SHARED}$. Although the area overhead grows as the size of the read access transistor increases, the area increase (12% and 7% for single-ended and differential, respectively) is not severe even at $N_{FIN,SHARED}$ of 50.

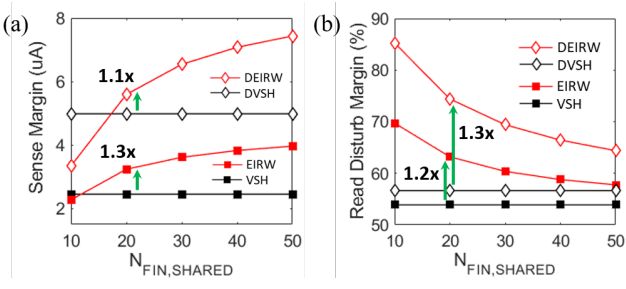

Fig. 9. (a) Sense margin (SM) and (b) read disturb margin (RDM) *versus* $N_{FIN,SHARED}$ of (D)EIRW-VSH MRAMs compared to (D)VSH-MRAMs.

### D. Array-level Analysis

In the previous section, device analysis shows that low $R_S$ in the read path of (D)EIRW-VSH MRAMs leads to 1.9X improvement in SM at RDM of ~80%. Hence, exchange-coupling-enabled EIRW is favorable to read performance in array designs including enhancement of SM and RDM, and reduction in read time and energy. However, this comes at the cost of higher write time and energy. In the following, we describe a comprehensive array-level analysis including 1) SM and area trade-off, and 2) time and energy analysis with various array sizes. Also, the results are compared with those of baseline (D)VSH-MRAM arrays.

*1) Read Performance*

The size of the shared read access transistor plays a key role in the performance of the proposed array design. For 20 fins of the shared read access transistor, 1.3X and 1.1X improvement in SM are observed for single-ended and differential arrays, respectively, compared to the (D)VSH-MRAM arrays (see Fig. 9(a)). Moreover, increasing the size of the shared read access transistor, which is proportional to $N_{FIN,SHARED}$, can further enhance SM up to 1.5X-1.6X at $N_{FIN,SHARED}$ = 50. It is important to note that there exists a trade-off between SM and area (Fig. 8(e) and Fig. 9(a)). Depending on the system requirements and target specification, $N_{FIN,SHARED}$ can be tuned for the required sensing robustness with a comparable or slight increase in area.

Second, the exchange-coupling between FLs helps in reducing read disturbance (see Section II for more details). Given that (D)EIRW-VSH MRAM exhibits higher $I_{CR}$ than (D)VSH-MRAM due to the exchange-coupling, (D)EIRW-

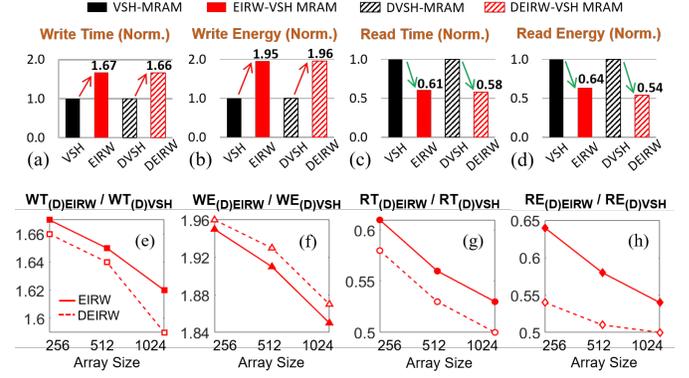

Fig. 10. (a) Write time (WT), (b) write energy (WE), (c) read time (RT), and (d) read energy (RE) of (D)EIRW-VSH MRAM arrays ($N_{FIN,SHARED}$ = 20). Values are normalized compared to (D)VSH-MRAMs. Normalized WT, WE, RT, RE are analyzed with various array sizes in (e-h).

VSH MRAM array shows 1.2X~1.3X higher RDM than (D)VSH-MRAM (Fig. 9(b)). Although increasing the size of the shared read access transistor leads to the reduction in RDM due to larger $I_{READ}$, RDM of (D)EIRW-VSH MRAM remains 1.1X higher than (D)VSH-MRAM even at $N_{FIN,SHARED}$ of 50.

*2) Write/Read Time and Energy*

Fig. 10(a)-(d) show write/read time and energy comparison of the proposed (D)EIRW-VSH MRAM arrays with the baseline (D)VSH-MRAM arrays. Read time and energy (RT and RE) are lowered in the proposed (D)EIRW-VSH MRAM arrays due to small $R_S$ in the read path and compact array design by virtue of the shared read access transistor. (D)EIRW-VSH MRAM arrays exhibit 39%-42% reduction in RT, and 36%-46% reduction in RE compared to the (D)VSH-MRAM based arrays. However, exchange-coupling in the proposed designs leads to 67% and 66% increase in write time (WT) for single-ended and differential arrays (array size = 256*256; $N_{FIN,SHARED}$ = 20). As the write energy (WE) is proportional to WT, the increased WT leads to 95% and 96% increase in WE for single-ended and differential arrays, respectively.

We further investigate the effect of array size on write and read performance (see Fig. 10(e)-(h)). Write/read time and energy are calculated with the various array sizes from 256*256 to 1024*1024, and normalized to (D)VSH-MRAM arrays. For write, increasing the array size reduces the WT and WE penalties in the proposed designs as the contribution of FL switching time diminishes with the increase in metal-line-charging time/energy. For read, the read time and energy improvements are further enhanced as array size is increased. Our results show RT and RE reduction of up to 47%-50% and 46%-50%, respectively, for a 1024*1024 array. This is because the effect of low $R_S$ in the proposed designs is manifested to a larger extent in larger arrays as the sense-line $RC$ delay becomes more dominant for larger array sizes (due to increase in the sense-line capacitance).

*3) Benchmarking with other Memories*

In Figure 11, we benchmark the memory metrics with SRAM, and emerging spin-based NVMs such as STT-MRAM [2], (D)GSH-MRAMs [7], [8], and (D)VSH-MRAMs [16]. Values of RT, RE, WT, WE, and area are obtained in 1024*1024 arrays, and normalized with respect to SRAM. SM is normalized to that of STT-MRAM as the SRAM employs voltage-based sensing while emerging spin-based NVMs use current-based sensing. For WE and RE, 30% cache utilization [28] is considered for all designs. Therefore, SRAM cache expends the leakage energy with the remaining 70% of idle time whereas emerging NVMs do not (i.e. zero leakage). Among the current-based designs, the proposed (D)EIRW-VSH MRAMs show the lowest RT and RE. RE of the proposed designs is comparable to that SRAM. Despite the write cost in the proposed MRAMs, WT and WE are lower than STT-MRAM. Compared to (D)GSH-MRAMs, WE is smaller in our designs. Furthermore, (D)EIRW-VSH MRAMs exhibit considerable area reduction compared to other memory solutions. However, these benefits come at the cost of lower SM (but also improved RT and RE) compared to STT- and GSH-MRAMs.

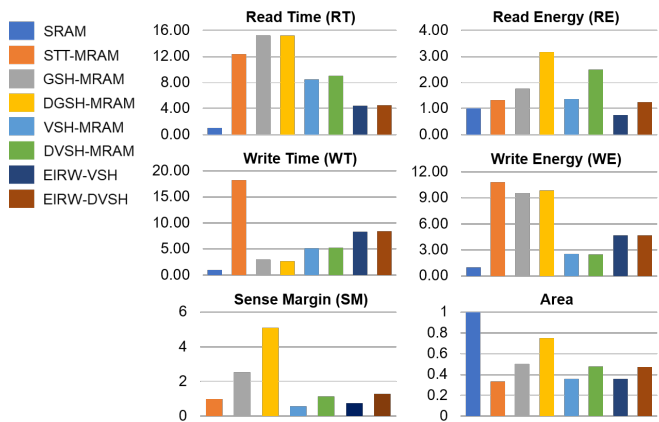

Fig. 11. Benchmarking of memory metrics with SRAM and emerging spin-based nonvolatile memories in array of 1024*1024. Values are normalized to SRAM except for sense margin, which is normalized to STT-MRAM.

## IV. Conclusion

The previously proposed VSH-MRAMs are known for compact layout and energy-efficient PMA based write without the assistance of external magnetic field, but the high series resistance in read path degrades the memory sense margin. To mitigate this issue, we propose VSH-based nonvolatile spintronic devices featuring electrical isolation of read and write paths (EIRW-VSH devices) enabled by exchange coupling between free layers. Since the large series resistance is averted in our proposed designs, we observe 1.9X improvement in sense margin at ~iso-read disturb margin of ~80% at the device level compared to the previously proposed VSH-devices. However, this comes at the cost of write efficiency as the exchange-coupling between free layers increases write latency by 1.7X in the proposed devices. We also explore array design for the proposed EIRW-VSH MRAMs employing a shared read access transistor for every word. With comparable area, the proposed memory arrays achieve 39%-42% and 36%-46% reduction in read time and energy, respectively, along with 1.1X-1.3X larger SM compared to VSH-MRAM based arrays, but at the cost of 1.7X and 2.0X write time and energy. We also observe that increasing the array size can further enhance the improvement in aforementioned read performance and mitigate the cost of write efficiency in our design. Thus, the proposed EIRW-VSH design is suitable for applications in which reads are more dominant than writes, while the baseline VSH design may be more suitable for applications which require frequent writes.

### Acknowledgment

The authors thank Prof. Zhihong Chen and Xiangkai Liu (Purdue) for their inputs on experimental characteristics of VSH-devices and Prof. Azad Naeemi (Georgia Tech) for his input regarding dipolar coupling on PMA magnets.

**Karam Cho** is a Ph.D. candidate (2018-present) in Electrical and Computer Engineering at Purdue University. She received her B.S. in Physics and M.S. in Electrical and Computer Engineering from University of Seoul, Seoul, Korea in 2015 and 2017, respectively. Her current research focus is on spin-based low-power device and circuit co-design utilizing 2D TMD materials. She is the recipient of Bilsland Dissertation Fellowship, awarded by Purdue University in 2022. She worked as a graduate research intern in the Components Research division of Intel Corporation, Oregon, USA in 2022.

**Sumeet Kumar Gupta** received the B. Tech. degree in Electrical Engineering from the Indian Institute of Technology, Delhi, India in 2006, and the M.S. and Ph.D. degrees in Electrical and Computer Engineering from Purdue University, West Lafayette IN in 2008 and 2012, respectively. Dr. Gupta is currently an Elmore Associate Professor of Electrical and Computer Engineering at Purdue University. His research interests include low power variation aware VLSI circuit design, neuromorphic computing, in-memory computing, nano-electronics and spintronics, device-circuit co-design and nano-scale device modeling and simulations.